\documentclass[notitlepage, floatfix, tightenlines, showpacs, showkeys, nofootinbib, aps]{revtex4-1}
\usepackage[centertags]{amsmath}
\usepackage{enumitem}

\newlist{STEP}{enumerate}{1}
\setlist[STEP]{label=\Roman*:}

\usepackage{bm}
\usepackage{amsmath}
\usepackage{amsfonts}
\usepackage{amssymb}
\usepackage[pdftex]{graphicx}
\usepackage{color}
\usepackage{balance}
\RequirePackage[colorlinks,citecolor=darkred,urlcolor=deepmagenta,linkcolor=darkblue,bookmarks=false]{hyperref}
\usepackage{subfig}
\usepackage[toc,title]{appendix}
\usepackage[misc]{ifsym}
\usepackage{adjustbox}
\usepackage{multirow}
\usepackage{makecell}

\allowdisplaybreaks
\def\bea{\begin{eqnarray}}
\def\eea{\end{eqnarray}}
\definecolor{darkblue}{rgb}{0.0, 0.0, 0.62}
\definecolor{deepmagenta}{rgb}{0.7, 0.01, 0.7}
\definecolor{darkred}{rgb}{0.55, 0.0, 0.0}
\begin{document}

\title{The dominating mode of two competing massive modes of quadratic gravity}

\author{Avijit Chowdhury}
\email[\Letter \hspace{0.1cm}]{chowdhury.avijit.001@gmail.com}%
\author{Semin Xavier}
\email[\Letter \hspace{0.1cm}]{seminxavier@iitb.ac.in}
\author{S. Shankaranarayanan}
\email[\Letter \hspace{0.1cm}]{shanki@phy.iitb.ac.in}
\affiliation{Department of Physics, Indian Institute of Technology Bombay, Mumbai 400076, India}

\begin{abstract}
    Over the last two decades, motivations for modified gravity have emerged from both theoretical and observational levels. $f(R)$ and Chern-Simons gravity have received more
attention as they are the simplest generalization. However, $f(R)$ and Chern-Simons gravity contain only an additional scalar (spin-0) degree of freedom and, as a result, do not include other modes of modified theories of gravity. In contrast, quadratic gravity (also referred to as Stelle gravity) is the most general second-order modification to 4-D general relativity and contains a massive spin-2 mode that is not present in $f(R)$ and Chern-Simons gravity. Using two different physical settings --- the gravitational wave energy-flux measured by the detectors and the backreaction of the emitted gravitational radiation on the spacetime of the remnant black hole --- we demonstrate that \emph{massive spin-2 mode carries more energy than the spin-0 mode}. Our analysis shows that the effects are pronounced for intermediate-mass black holes, which are prime targets for LISA.
\end{abstract}

\maketitle
\balance
\section*{Introduction}
In quantum gravity, it is impossible to localize an event with a precision smaller than the Planck length ($\ell_{\rm Pl}$)~\cite{1994-Ng.VanDam-MPLA,1995-Garay-IJMPA,1999-Adler.Santiago.MPLA,2004-Calmet.etal-PRL}. From the hoop conjecture and the uncertainty principle, we can deduce the existence of a Planck-sized ball~\cite{1972-Thorne-Conf}. 
Thus, the operational significance of the concept of space-time points is lost~\cite{1964-deWitt-PRL,1981-deWitt-PRL,1993-Kuo.Ford-PRD,1997-Padmanabhan-PRL}. Most approaches to quantum gravity incorporate the Planck length by considering extended structures, rather than point particles, as fundamental blocks~\cite{Hossenfelder:2012jw,2013-Amelino-Camelia-LRR}. The Generalized Uncertainty Principle (GUP) is a phenomenological approach that introduces modifications to the Heisenberg uncertainty principle in the ultraviolet regime and studies its consequences~\cite{Kempf1995-yi,Pedram2010-nc}. Kempf et al. considered the following simplest modification to the canonical commutation relation~\cite{Kempf1995-yi}:
 $$
 [\hat{x}, \hat{p}]=i \hbar\left(1+\beta p^{2}\right) \quad \mbox{leading to} \quad 
 \Delta_{x} \geq \hbar \sqrt{\beta} \, .
 $$
$\beta$ is a parameter characterizing the GUP
whose value needs to be fixed by observations. Since the above GUP is non-relativistic, it is impossible to compute GUP corrections in relativistic field theories. Recently, Todorinov et al.~\cite{Todorinov:2018arx} extended the above GUP to a generic class of covariant GUPs: 
\begin{equation}
    \label{RGUP}
    \begin{split}
             [x^{\mu},p^{\nu}]=i\hbar \eta^{\mu\nu}(1 - \gamma p^{\sigma}p_{\sigma})-2i \hbar \gamma p^{\mu}p^{\nu},~\gamma= \gamma_0\ell_{\text{Pl}}^2
    \end{split}
\end{equation}
and studied the phenomenological features of such GUPs for scalar, spinor, and vector fields~\cite{Bosso2020-dv,Bosso2020-fz}. 
$\gamma_0$ is a positive numerical constant to be fixed by observations. There exists a correspondence between the GUP-modified dynamics of a massless spin-2 field and quadratic gravity (QG) with suitably constrained parameters\cite{nenmeli2021PLB},. Specifically, the authors showed that the 4-D gravity action~\cite{stelle1977PRD,stelle1978GERG} 
\begin{equation}
\label{eq:action}
S_{\rm QG}=  \int {d^{4}x \sqrt{-g}\left[\frac{1}{2\kappa^{2}}R-\alpha R_{\mu\nu}R^{\mu\nu}+\beta R^{2}\right]},
\end{equation} with $\kappa^{2}= \frac{8\pi G}{c^4}$
is a classical manifestation of the imposition of a momentum cutoff at the quantum gravity level when $\alpha=2\beta=\gamma/\kappa^{2}$. QG or Stelle gravity --- unlike $f(R)$ --- has extra massive spin-2 and spin-0 modes~\cite{stelle1978GERG}. Intriguingly, for this class of Stelle theories, the masses of the spin-0 and spin-2 modes  \emph{coincide} $(1/\sqrt{2 \gamma})$. 
The issue of unitarity in these theories is not completely settled yet~\cite{salam1978PRD,fradkin1981NPB, Antoniadis1986-hj,johnston1987NPB} \and the massive spin-2 mode may be of ghost nature. See, for instance, the fakeon approach for the same~\cite{anselmi2017JHEP,anselmi2020JHEP}.
Given that the Stelle gravity is the most general QG in 4-D, we ask the following questions: For identical masses, what is the role of the massive spin-2 mode compared to the spin-0 mode? What are the leading order corrections of QG compared to $f(R)$ gravity? This work addresses these questions by evaluating the corrections to the radiation in QG.

Over the last two decades, motivations for modified gravity have emerged from both theoretical and observational levels~\cite{Alexander:2009tp,DeFelice:2010aj,2014-Will-LRR,Nojiri:2017ncd,2022-Shanki.Joseph-GRG}. Since $f(R)$ and Chern-Simons gravity are the simplest generalization, they have received more attention. However, $f(R)$ and Chern-Simons gravity contain \emph{only the additional scalar degree of freedom} and hence do not contain other modes of modified theories of gravity~\cite{odintsov2021PDU}. We explicitly show that the massive spin-2 mode carries \emph{more energy than} the spin-0 mode using two different physical settings: {the energy flux of the gravitational waves, measured by an asymptotically placed gravitational wave detector, and the backreaction of the emitted gravitational radiation on the spacetime of the remnant black hole}. Thus, we show that the leading order gravitational radiation correction in QG is linear in the coupling constant compared to $f(R)$, where the corrections are quadratic in flat space-time~\cite{berry2011PRD}.

To show this explicitly, we start with the GUP-inspired version of QG action \eqref{eq:action}, where~\cite{nenmeli2021PLB}
\[
\alpha = {\gamma}/{\kappa^2}; \quad 
\beta = {\gamma}/{2 \kappa^2} \, .
\]
Varying action~\eqref{eq:action} with respect to the metric leads to: 
{ 
\begin{equation}\label{eq:stelleEOM}
\begin{split}
{\mathcal{G}}_{\mu\nu} \equiv &G_{\mu\nu}- 2 \gamma \Box G_{\mu \nu}-4 \gamma R^{\rho\sigma}R_{\mu\rho\nu\sigma} +2 \gamma R R_{\mu\nu} +\gamma g_{\mu\nu}\left[R^{\rho\sigma}R_{\rho\sigma}-\frac{R^{2}}{2}\right] = 0 \, .
\end{split}
\end{equation}
}
Linearizing the field equation~\eqref{eq:stelleEOM} about the Minkowski space-time ($\eta_{\mu\nu}$), we get, 
\begin{equation}
g_{\mu\nu}=\eta_{\mu\nu}+ \epsilon h_{\mu\nu}~,~ g^{\mu\nu}=\eta^{\mu\nu}- \epsilon h^{\mu\nu}~\mbox{and}~  h=\eta^{\mu \nu} h_{\mu \nu}, 
\end{equation}
where $h_{\mu\nu}$ is the metric perturbation, and $\epsilon$ is a book-keeping parameter. This leads to the following linearized equations:
\begin{equation}\label{eq:linEOM_flat}
\mathcal{\delta G}_{\mu \nu}\equiv \left(1- 2 \gamma \bar\Box\right)\left(\delta R_{\mu \nu}-\frac{1}{2} \eta_{\mu \nu} \delta R\right)=0~,
\end{equation} 
where $\bar\Box \equiv \eta^{\mu\nu}\partial_{\mu}\partial_{\nu}$ and
\begin{eqnarray}
\delta R_{\mu \nu}&=& \frac{1}{2}\left( \partial_\mu \partial_\rho h^\rho_\nu + \partial_\nu \partial_\rho h^\rho_\mu -\partial_\mu \partial_\nu h - \bar{\Box} h_{\mu \nu}\right),\label{eq:ricci}\\
\delta R &=& \eta^{\mu \nu}\delta R_{\mu \nu} = \left(\partial_\mu \partial_\rho h^{\rho\mu}-\bar{\Box} h\right) \, . \label{eq:ricciscalar}
\end{eqnarray}
As expected in higher-derivative gravity theories, Eq. \eqref{eq:linEOM_flat} contains fourth-order derivatives whose trace is given by: 
\begin{equation}\label{eq:lintrcEOM_flatb}
\bar{\Box}\delta R -  \delta R/{2 \gamma}=0.
\end{equation}
To separate the contribution of the different spin-components from linearized field equation~\eqref{eq:linEOM_flat}, we use the following ansatz for the metric perturbations {(See Appendix~\ref{app:Msv-bigrav} for a detailed calculation in Ricci-flat spacetime)}:
\begin{equation}\label{eq:ansatz}
h_{\mu\nu}=\left[\psi_{\mu \nu}- \frac{\eta_{\mu\nu}}{2} \psi \right]+ \left[C_1+ \frac{C_2}{4}\right] R^{(1)} \eta_{\mu\nu} - C_2 \hat{R}^{(1)}_{\mu\nu}~,
\end{equation}
where $\psi= \eta^{\alpha \beta} \psi_{\alpha \beta}$ and $C_1, C_2$ are arbitrary constants, and $\hat{R}^{(1)}_{\mu\nu}=R^{(1)}_{\mu\nu}-\frac{1}{4} \eta^{\mu\nu} R^{(1)}$
is the traceless part of $ R^{(1)}_{\mu\nu}$. To our knowledge, this is the first time an explicit separation of metric fluctuations is used in this context, and we would like to emphasize the following points: First, in general relativity (GR), the constants $C_1$ and $C_2$ vanish because the linearized field equations only contain massless spin-2 (graviton) mode.
Thus, in GR, $\psi_{\mu \nu}$ reduces to the familiar trace-reversed metric perturbations. Second, in the case of Starobinsky model, the field equations contain additional contribution from the massive spin-0 mode $(R^{(1)})$ only, hence $C_1=-6\gamma$ and $C_2=0$~\cite{berry2011PRD,2017-Soham.Shanki-PRD}.
{Third, the existence of the massive spin-2 ghost degrees of freedom may suggest a potential pathology due to Ostrogradsky instability~\cite{stelle1977PRD},  in the present context it is possible to avoid such a pathology by treating the massive and massless spin-2 modes as a single structure\cite{alves2022arxiv}.}
Lastly, like in GR, we use the de-Donder gauge on the residual massless graviton mode ($\psi_{\mu\nu}$) to restrict the gauge freedom:
\begin{equation}\label{eq:lorenz_flat}
\partial^\mu \psi_{\mu\nu}=0 \, .
\end{equation}
Substituting Eq.~\eqref{eq:ansatz} in Eqs.~(\ref{eq:ricci},\ref{eq:ricciscalar}) and using the gauge conditions~\eqref{eq:lorenz_flat}, we get: 
\begin{equation}\label{eq:GW_flat}
2 \, \mathcal{\delta G}_{\mu \nu}=-\bar{\Box}\psi_{\mu\nu} ~.
\end{equation}
where we set $C_1=-2\gamma$ and $C_2=4\gamma$. Taking cognizance of Eqs.~(\ref{eq:linEOM_flat},~\ref{eq:GW_flat}) and \eqref{eq:lintrcEOM_flatb}, we get the propagation equation for the graviton, the massive spin-2 and spin-0 modes as, 
\begin{eqnarray}
\bar{\Box}\psi_{\mu\nu}= 0~,\label{eq:M0S2_flat}\\
\bar{\Box}\hat{R}^{(1)}_{\mu \nu} -  \hat{R}^{(1)}_{\mu \nu}/{2 \gamma} = 0~, \label{eq:MS2_flat}\\
\bar{\Box}R^{(1)} -  R^{(1)}/{2 \gamma}=0~
\label{eq:lintrcEOM_flat}
\end{eqnarray}
(See Appendix \ref{app:EOM-generic} for a detailed derivation of the propagation equations in a Ricci-flat background).
Thus, the theory~\eqref{eq:action} is described by a massless graviton ($\psi_{\mu\nu}$); a massive spin-2 ($\hat{R}^{(1)}_{\mu \nu}$) and a massive spin-0 (${R}^{(1)}$)~\cite{stelle1977PRD, stelle1978GERG, nenmeli2021PLB,tachinami2021PRD}
(The massive spin-2 mode can be further decomposed into two tensor (helicity-2) modes, two vector (helicity-1) modes, and scalar (helicity-0) mode, whereas the massless spin-2 graviton gives rise two tensor (helicity-2) modes and the massive spin-0 particle has one scalar (helicity-0) mode. The different helicity modes can, in principle, be mapped to the six metric degrees of freedom denoting the six polarization modes of the gravitational waves~\cite{tachinami2021PRD, eardley1973PRD, will2014LRR,lee1974PRD}). 
The positivity of the parameter $\gamma$ ensures that the massive modes are non-tachyonic.
Repeating the analysis for the Ricci-flat background by  
using the following ansatz:
\begin{equation}\label{eq:ansatzRicciflat}
h_{\mu\nu}=\left(\psi_{\mu \nu}- \frac{1}{2} \psi \bar{g}_{\mu\nu}\right)- \bar{g}_{\mu\nu} \gamma R^{(1)} - 4 \gamma \hat{R}^{(1)}_{\mu\nu}~,
\end{equation}
the equations of motion for the massless \eqref{eq:M0S2_flat} and massive spin-2 modes \eqref{eq:MS2_flat} in the transverse traceless gauge $\left(\bar\nabla^\mu \psi_{\mu\nu}=0, ~\bar{g}^{\mu\nu} \psi_{\mu\nu}=0\right)$ are:
\begin{eqnarray}
\bar{\Box}\psi_{\mu\nu}+ 2 \bar{R}_{\mu\alpha\nu\beta} \psi^{\alpha\beta} = 0~, \label{eq:Rflatboxpsi}\\
\bar{\Box}\hat{R}^{(1)}_{\mu \nu} +2 \bar{R}_{\mu\alpha \nu\beta} \hat{R}^{(1) \alpha \beta} - \hat{R}^{(1)}_{\mu \nu}/{(2 \gamma)} =0,\label{eq:RflatboxRic}
\end{eqnarray}
where $\bar{g}_{\mu\nu}$ is the background metric, $\bar{R}_{\mu\alpha \nu\beta}$ is the background Riemann tensor, $\bar\Box\equiv \Bar{\nabla}_\sigma \Bar{\nabla}^\sigma$ with $\bar{\nabla}^\sigma$ being the covariant derivative due to the background spacetime and
the \emph{traceless} tensor $\hat{R}^{(1)}_{\mu\nu}$ is
\begin{equation}\label{eq:spin-2tracls}
\hat{R}^{(1)}_{\mu\nu}=R^{(1)}_{\mu\nu}- \Bar{g}^{\mu\nu} R^{(1)}/4~.
\end{equation}
The propagation of the massive spin-0 mode is still governed by Eq.~\eqref{eq:lintrcEOM_flat}, with the above-defined D'Alembertian operator. The mass degeneracy between the two massive (spin-0 and spin-2) modes demonstrates that they are not completely independent and are related by linearized Bianchi identities:
\begin{equation}\label{eq:pertbian}
4 \, \bar\nabla_\mu \hat{R}^{(1) \mu \nu}-\ \bar g^{\mu\nu} \bar\nabla_\mu R^{(1)} 
= 0 \, .
\end{equation}

Having separated the metric fluctuations into massless and massive modes in Ricci flat background, we now evaluate the energy and momentum carried by the gravitational waves in degenerate-Stelle gravity. 
To go about that, we expand the field equation to 
second-order in $\epsilon$:
\begin{equation}\label{eq:modG-exp}
\mathcal{G}_{\mu\nu}=\bar{\mathcal{G}}_{\mu\nu}+\epsilon \mathcal{\delta G}_{\mu\nu}+\epsilon^2 \delta^2\mathcal{G}_{\mu\nu}=0,
\end{equation}
where $\bar{\mathcal{G}}_{\mu\nu}$ represents the background quantity, $\delta \mathcal{G}_{\mu\nu}$ are linear in perturbations ($h_{\mu\nu}$) and $\delta^2 \mathcal{G}_{\mu\nu}$ are quadratic in $h_{\mu\nu}$. Through second-order perturbations, Isaacson established a procedure to obtain an effective stress-energy tensor for gravitational radiation~\cite{isaacson1968PR1, isaacson1968PR}. Specifically, the effective stress-energy tensor of the emitted gravitational waves is obtained by averaging over a length-scale $l$ such that $\lambdabar \ll l \ll \Lambda$, where $\lambdabar$ is the wavelength of the fluctuations and $\Lambda$ is the system size. The short wavelength components will be averaged out, yielding a gauge-invariant measure of the effective gravitational wave (GW) stress-energy tensor~\cite{brill1964PR}:
\begin{equation}
t_{\mu\nu}^{\rm GW} = \frac{1}{\kappa^2}\left<\Bar{\mathcal{G}}_{\mu\nu} \right>=- \frac{1}{\kappa^2} \left< \delta^2\mathcal{G}_{\mu\nu} \right>.
\end{equation}

In the Ricci-flat background, we get
\begin{widetext}
\begin{equation}
\label{eq:teff_full_trcls}
\begin{split}
    t_{\mu\nu}^{\rm GW} = &\frac{1}{\kappa^2}\Bigl[\left<\frac{1}{4}\bar\nabla_\mu \psi^{\rho \sigma}\bar\nabla_\nu \psi_{\rho \sigma}
-  \gamma \left(\bar\nabla_\mu \psi^{\rho \sigma} \bar\nabla_\nu\hat{R}^{(1)}_{\rho \sigma} + \bar\nabla_\nu \psi^{\rho \sigma} \bar\nabla_\mu\hat{R}^{(1)}_{\rho \sigma}\right)\right.\\
&\left.-\gamma^2 \left(4 \bar\nabla_\mu \hat{R}^{(1)\rho \sigma} \bar\nabla_\nu \hat{R}^{(1)}_{\rho \sigma}- \bar\nabla_\mu R^{(1)}\bar\nabla_\nu R^{(1)} \right)\right>
+\left<\mathcal{A}_{\mu\nu}+\gamma \mathcal{B}_{\mu\nu}+\gamma^2 \mathcal{C}_{\mu\nu}+\gamma^3 \mathcal{D}_{\mu\nu}\right>\Bigr],
\end{split}
\end{equation}
\end{widetext}
where $\mathcal{A}_{\mu\nu},\mathcal{B}_{\mu\nu},\mathcal{C}_{\mu\nu},\mathcal{D}_{\mu\nu}$ are related to the background Riemann tensor (and are explicitly given in Appendix~\ref{app:GWSET}). This is the first key result of this work, regarding which we would highlight the following points: 
First, in the Minkowski limit (as in the case of GW detectors), $\mathcal{A}_{\mu\nu},\mathcal{B}_{\mu\nu},\mathcal{C}_{\mu\nu},\mathcal{D}_{\mu\nu}$ vanish and 
$t_{\mu\nu}^{\rm GW}$ is proportional to partial derivatives of $\psi^{\rho\sigma}$, $\hat{R}^{(1)}_{\rho \sigma}$ and $R^{(1)}$. 
Second, the first term within the triangular bracket gives the dominant contribution --- the contribution of the graviton mode as in GR~\cite{isaacson1968PR1}. However, the \emph{crucial difference is the dominant contribution} of the massive spin-2 mode. The massive spin-2 mode contribution is proportional to $\gamma$, whereas the massive spin-0 mode contribution is proportional to $\gamma^2$. Thus, the above expression explicitly shows that the massive spin-2 mode carries more energy than the massive spin-0 mode. 
Third, while we have used Isaacson's approach to obtain the stress-tensor, other approaches also give similar results\cite{saffer2018CQG}.
Fourth, the leading order contribution of the massive spin-2 mode is opposite to that of the graviton mode.
Lastly, the corrections by QG to GR are much larger than $f(R)$ gravity~\cite{berry2011PRD,bhattacharyya2018EPJC}. Consequently, our study demonstrates that the $f(R)$ theories \emph{overlook crucial information} concerning the massive spin-2 mode.
In what follows, we use the  GW stress-energy tensor \eqref{eq:teff_full_trcls} to examine the effect of the massive spin-2 mode under two distinct physical settings.
First, we investigate the energy flux of GWs as measured by the \emph{GW detectors at asymptotic infinity}. We then estimate the change in the spherically symmetric metric caused by the backreaction of the \emph{emitted GWs near the horizon.}

\section*{Gravitational wave energy flux:} The energy of the gravitational wave within a spatial volume $V$ is~\cite{MTW, Maggiore}
\begin{equation}
E_V=\int_V d^3x~  t^{00}_{\rm GW}.
\end{equation}
Using the stress-tensor conservation equation ($\partial_\mu t^{\mu\nu}_{\rm GW} = 0$) for the far-away observer, 
the power carried by the gravitational waves is~\cite{saffer2018CQG, Maggiore}
\begin{equation}\label{eq:GW_power}
\frac{dE}{d (ct)}=- \oint_S t^{0 i}_{\rm GW} \,  n_i dA
\end{equation}
where $S$ is the surface enclosing the volume $V$ and $n_i$ is the unit outward normal to $S$.
The negative sign signifies that an outward propagating gravitational wave carries away energy from the source. Thus, plugging the Minkowski-limit of Eq.~\eqref{eq:teff_full_trcls} in Eq.~\eqref{eq:GW_power} and considering $S$ to be a spherical surface at a large distance from the source, the gravitational wave energy flux passing through the detector is:

\begin{equation}\label{eq:power/solidangle}
\begin{split}
    \frac{dE}{dt dA}=\frac{c}{\kappa^2}\left<\frac{1}{4}\partial^0 \psi^{\rho \sigma} \partial_r \psi_{\rho \sigma}\right.\left.-\gamma \left[\partial^0 \psi^{\rho \sigma} \partial_r\hat{R}^{(1)}_{\rho \sigma} + \partial^0\hat{R}^{(1)}_{\rho \sigma}  \partial_r \psi^{\rho \sigma}\right]\right.\left.-\gamma^2 \left[  4 \partial^0 \hat{R}^{(1)\rho \sigma} {\partial_r} \hat{R}^{(1)\rho \sigma}-  \partial^0 R^{(1)}{\partial_r} R^{(1)} \right]  \right>
\end{split}
\end{equation}

To make the calculations transparent, we assume that all the three (graviton, massive spin-2, and spin-0) modes of the following form:
\begin{equation}
\label{eq:planewave}
{ 
\begin{split}
    & \psi_{\mu\nu} =\zeta_{\mu\nu}  e^{i k_\rho x^\rho}+\zeta^*_{\mu\nu}  e^{-i k_\rho x^\rho} ;~ R^{(1)} =\phi  e^{i q_\rho x^\rho}+\phi^{*}  e^{-i q_\rho x^\rho} ; \\
& \hat{R}^{(1)}_{\mu\nu} =\theta_{\mu\nu}  e^{i w_\rho x^\rho}+\theta^{*}_{\mu\nu}  e^{-i w_\rho x^\rho} 
\end{split}
}
\end{equation}
where $\zeta_{\mu\nu}, \theta_{\mu\nu}$ and $\phi$ depend on $r$ and $t$,
\begin{equation}
|\mathbf{k}| =\omega/c~~;~~
|\mathbf{w (q)}|=\sqrt{\omega^2/c^2 - 1/2 \gamma}
\end{equation}
with $k^0= w^0 = q^0 = \omega/c$,  $\mathbf{k}=\left(k^i\right)$,  $\mathbf{w}=(w^i)$, and $\mathbf{q}= (q^i) $. $\omega > c/\sqrt{2\gamma}$ and $q > c/\sqrt{2\gamma}$ corresponds to  
oscillatory solution for $\hat{R}^{(1)}_{\mu\nu}$ and $R^{(1)}$, respectively.  
A wave propagating radially outward $(\Phi_{\mu\nu})$ at large distances from the source can be represented to fall-off radially~\cite{Maggiore,poisson}:
\begin{equation}
\Phi_{\mu\nu}=\frac{1}{r} \chi_{\mu\nu}(t_r); 
\end{equation}
where $\chi_{\mu\nu}(t_r)$ is an arbitrary function of the retarded time $t_r=t-r/v$, and $v=c/\sqrt{1-\frac{1}{2 \gamma}\left(\frac{c}{\omega}\right)^2}$ is the speed of the massive modes. Thus,
{ 
\begin{equation}
\begin{split}
\label{eq:partial_r-t}
\frac{\partial}{\partial r} \chi_{\mu\nu}  (t-r/v) =-\frac{1}{v}\frac{\partial}{\partial t}\chi_{\mu\nu} (t-r/v) \\
\Longrightarrow \frac{\partial}{\partial r} \zeta_{\mu\nu}=-\frac{c}{v}\partial_0 \zeta_{\mu\nu}+O\left(\frac{1}{r^2}\right)
\end{split}
\end{equation}
}
Substituting Eq.~\eqref{eq:planewave} in Eq.~\eqref{eq:power/solidangle} and using Eq.~\eqref{eq:partial_r-t} in the resultant expression
leads to the energy flux on the GW detectors:
\begin{equation}\label{eq:power/solidangle-final}
\begin{split}
\frac{dE}{dt dA}=\frac{c^3}{8\pi G}\left<\frac{1}{4}\dot{\psi}^{\rho \sigma}\dot{\psi}_{\rho \sigma} 
-  \gamma \left(1+\frac{c}{v}\right) \dot{\psi}^{\rho \sigma} \dot{\hat{R}}^{(1)}_{\rho \sigma}\right.
\left.-\gamma^2  \frac{c}{v}  \left( 4\dot{\hat{R}}^{(1)\rho\sigma} \dot{\hat{R}}^{(1)}_{\rho\sigma}- \left(\dot{{R}}^{(1)}\right)^2 \right) \right>~,
\end{split}
\end{equation}
where dot denotes derivative w.r.t $t$.
Here again, we note that the dominant contribution comes from the graviton mode with leading order corrections $(\gamma)$ from the massive spin-2 mode; the massive spin-0 mode contributes only in the second order. Hence, as expected, the 
measured energy energy-flux in the case of QG is lower than that predicted by GR. In other words, the massive spin-2 mode carries more energy than the spin-0 mode. Since, this analysis is for the Minkowski background, $\gamma {\cal B}_{\mu\nu}$ (in Eq. \ref{eq:teff_full_trcls}) is zero. However, in the case of curved geometry, $\gamma {\cal B}_{\mu\nu}$ contribution might be significantly larger than the linear order term in the above expression.

In the case of GW detectors, the average $<\ldots>$ is purely a temporal average~\cite{Maggiore}, and the total energy flowing through the unit area of the detector is:
\begin{equation}
\begin{split}
    \frac{dE}{dA}=\frac{c^3}{8\pi G}\int_{-\infty}^{\infty} dt \left[\frac{1}{4}\dot{\psi}^{\rho \sigma}\dot{\psi}_{\rho \sigma} 
-  \gamma \left(1+\frac{c}{v}\right) \dot{\psi}^{\rho \sigma} \dot{\hat{R}}^{(1)}_{\rho \sigma}\right.
\left.-\gamma^2  \frac{c}{v}  \left( 4\dot{\hat{R}}^{(1)\rho\sigma} \dot{\hat{R}}^{(1)}_{\rho\sigma}- \left(\dot{{R}}^{(1)}\right)^2 \right) \right]~.
\end{split}
\end{equation}
Note that the above analysis is strictly valid for  $\omega > c/\sqrt{2\gamma}$ and $q > c/\sqrt{2\gamma}$ corresponding to  
oscillatory solutions for the two massive modes. In the second physical setting, we will relax this condition and obtain the contribution of the massive spin-2 mode.


\section*{Backreaction of the emitted gravitational waves}
To study the backreaction of the emitted GWs on the background black hole space-time, we assume that the background space-time is spherically symmetric Ricci flat and is an exact solution of the QG action \eqref{eq:action}.
Though Schwarzschild solution in Stelle gravity is known to suffer from  $\ell =0$ mode instability (Gregory-Laflamme instability) when the spin-0 mode is non-propagating~\cite{myung2013PRD, myung2018PLB, brito2013PRD,lu2017PRD}, in the present work, we consider the spin-0 mode to be dynamical and concentrate on the $\ell=2$ modes.
The backreaction on the background space-time is obtained by 
evaluating the GW stress-energy tensor ($t_{\mu\nu}^{\rm GW}$) 
w.r.t the Schwarzschild metric. Having obtained this, we then solve the effective Einstein's equations:
\begin{equation}\label{eq:modEin}
{G_{\mu}^{\nu}}^{\rm mod}=\kappa^2~{t_{\mu}^{\nu}}^{\rm GW} \, ,
\end{equation}
where the Einstein tensor ${G_{\mu}^{\nu}}^{\rm mod}$ is evaluated for spherically symmetric space-time in dimensionless, Eddington-Finkelstein (EF) coordinates~\cite{MTW}:
\begin{equation}\label{eq:modsch}
ds^2=-e^{2\nu}dV^2+2e^{\nu +\lambda}dV d\rho +\rho^2 d\Omega^2 ,
\end{equation}
$\nu \equiv \nu\left(V,\rho\right)$ and $\lambda \equiv \lambda\left(V,\rho\right)$ encode the corrections from the emitted GWs, and $d\Omega^2$ is 
the metric on unit 2-sphere. Note that $V$ and $\rho$ are dimensionless like $\theta$ and $\phi$.  
Regarding Eqs. (\ref{eq:modEin}, \ref{eq:modsch}), the following points are in order: First, the ansatz \eqref{eq:ansatzRicciflat} assumes that all the metric components are dimensionless. Hence, we have rescaled all the coordinates to be dimensionless.
Second, $V=constant$ hypersurfaces represent the ingoing null geodesics. As mentioned, the background metric is assumed to be Schwarzschild black hole; hence, $e^{2\nu}=e^{-2\lambda}=1-2M_0/\rho$ where $M_0$ is the dimensionless mass parameter (setting $c= G = 1$). For $M(V)$, the above metric gives the Vaidya line element. 
Third, the ingoing EF coordinates are smooth across the horizon for ingoing null geodesics and are suitable for analyzing the gravitational waveform close to the horizon~\cite{philipp2015PRD, philipp2015IJMPD} as well as the shift in the horizon radius due to the backreaction. Finally, since $t_{\mu\nu}^{\rm GW}$ contains the contributions of graviton and massive modes, the LHS of Eq.~\eqref{eq:modEin} only contains the Einstein gravity. Note that the radiation from a remnant black hole decreases its energy content, inducing a change in the metric.

Since the dominant contributions to the GW stress-energy tensor~\eqref{eq:teff_full_trcls} come \emph{only from the spin-2 mode}, to evaluate their effects on the metric, we use the following ansatz:
\begin{equation}\label{eq:psisplit}
\psi^{\mu\nu} = o^{\mu\nu} \psi(V,\rho) Y^m_\ell(\theta, \phi);~~
\hat{R}^{(1) \mu\nu}= \iota^{\mu\nu} P(V,\rho) Y^m_\ell(\theta, \phi),
\end{equation}
where $o^{\mu\nu}$, $\iota^{\mu\nu}$ are constant, traceless (polarization) tensors, $\psi(V,\rho)$, $P(V,\rho)$ are scalar functions and $Y^m_l(\theta, \phi)$ are spherical harmonics. 
The above assumptions 
essentially replaces the spin-2 modes~$\psi^{\mu\nu},~\hat{R}^{(1) \mu\nu}$ by scalar functions, where we ignored the non-linear transformation among the components of the individual spin-2 modes.
To obtain the leading order corrections, we concentrate on the backreaction effect due to the $\ell=2$ and $m=0$ mode of the gravitational waveform and obtain the average contribution over the entire solid angle. Note that the $(\ell=2,m=0)$ mode contributes to the non-linear memory of the GWs, which is otherwise difficult to observe in ground-based GW detectors~\cite{favata2010CQG,mehta2017PRD} and nontrivial to extract in numerical relativity simulations. We can trivially extend the analysis to other modes.

We assume the modified black hole to be described by the generalized spherically symmetric metric ansatz proposed by Johannsen and Psaltis (JP)~\cite{johannsen2011PRD,rezzolla2014PRD}. In the (dimensionless) EF coordinates \eqref{eq:modsch}, we have:
\begin{equation}
\begin{split}
e^{2\nu}=f(\rho)\left[1- \frac{2\tilde{M}}{\rho} \right];\quad e^{2\lambda}=\frac{f(\rho)}{1-2\tilde{M}/\rho};\quad
f(\rho)=\sum_{n=0}^{\infty} \tilde{\epsilon}_n \left(\frac{\tilde{M}}{\rho}\right)^n.
\end{split}
\end{equation}
where $\tilde{\epsilon}_0=1$ and the first few coefficients of the expansion can be constrained from the PPN-like parameters~\cite{johannsen2011PRD}. In the limit of $\tilde{\epsilon}_n=0,~(n>0)$, JP metric reduces to the Schwarzschild metric. The event horizon of the corresponding black hole is at $\rho=2\tilde{M}$ and the (dimensionless) ADM mass is $M_{\rm ADM} =\tilde{M}(1-\tilde{\epsilon}_1 /2)$~\cite{rezzolla2014PRD}. As mentioned earlier, the remnant black hole decreases its energy content, inducing a change in the metric mass-function from the initial, dimensionless Schwarzschild value $(M_0)$~\cite{abramo2001GERG, Kimura2021PTEP}:
\begin{equation}
\tilde{M} = M_0+\Delta_M(V)
\end{equation}

Assuming the mode functions to be \emph{regular and slowly varying} in $\rho$ close to horizon $(\Delta=\rho-2M_0<<1)$~\cite{philipp2015PRD, philipp2015IJMPD, rajesh2020PRD}, and expanding both sides of Eq.~\eqref{eq:modEin} for the $\rho-V$ component, we get:
\begin{equation}
\label{eq:JP}
\sum_{n=0}^\infty \frac{2^{n-1}}{\tilde{\epsilon}_n}~~ \frac{\Delta_M'(V)}{M_0^2} \approx \frac{ M_0^4\psi'^2}{21\pi} \sum_{\substack{(k,l)=\\ \{(0,0),(2,3)\}}}C^{kl}\Biggl[1-\frac{\iota^{kl}\gamma P'}{o^{kl} \psi'} \Biggr]
\end{equation}
where $\{C^{00},C^{23}\}=\{29(o^{00})^2,20(o^{23})^2\} $ and  $'$ denotes derivative with respect to $V$. The 
expression is evaluated to the leading order in $\Delta$ and $M_0$.
Integrating from some initial $V=V_0$, when $\Delta_M(V = V_0)=0$ to some final $V=V_1$, we get,
\begin{equation}\label{eq:vaidyaM}
\Delta_M(V_1) =   A\int_{V=V_0}^{V1} dV \sum_{\substack{(k,l)=\\ \{(0,0),(2,3)\}}} C^{kl}\Biggl[1-\frac{\iota^{kl}\gamma P'}{o^{kl} \psi'} \Biggr],
\end{equation}
where 
\begin{equation}
    A= \frac{ M_0^6\psi'^2}{21\pi} \sum_{n=0}^{\infty}\frac{\tilde{\epsilon}_n}{2^{n-1}}
\end{equation}
This is the third key result regarding which we would like to stress the following: 
First, the massive and massless spin-2 modes contribute oppositely to the change in the mass function, hence the shift in the horizon radius. Second, since $\Delta_M$ is proportional to the sixth power of $M_0$, it implies that the larger the mass of the perturbed black hole, the larger the corrections to the change in the mass. This assumes particular importance for intermediate-mass black holes which are prime targets for LISA.
Rezzolla and Zhidenko~\cite{rezzolla2014PRD} proposed an improved parametric metric in which, in the near-horizon limit, the Taylor expansion is replaced by an expansion in continued fractions (CF). The Rezzolla-Zhidenko metric has better convergence properties and can effectively reproduce any  known solution even in scenarios where the JP parametrization fails. The CF expansion coefficients in the Rezzolla-Zhidenko metric can be expressed in terms of the JP parameters; hence it is possible to extend our results to the Rezzolla-Zhidenko metric~\cite{rezzolla2014PRD}.


\section*{Summary and Discussions}
In this work, we examined gravitational radiation in QG. We explicitly decomposed the GWs in Stelle gravity into massless and massive spin-2 and spin-0 modes. We demonstrated that the dominant contribution to the GW stress-energy tensor comes from the graviton mode, with leading order corrections coming from the massive spin-2 mode, which is absent in $f(R)$ gravity theories. 
We can ascertain this because the massive spin-2 and spin-0 have the same mass in the case of GUP-inspired Stelle gravity.
In contrast, the effects of the massive spin-2 mode are the inverse of those of the GR mode. In the context of GW detectors, this results in a decreased energy flux measurement. This result is consistent with a recent finding that this Stelle gravity model reduces the amplitude of primordial gravitational waves produced by Starobinsky inflation~\cite{2022-Das.etal}.

We also provided an estimate of the backreaction effect of the GW emission on the background spacetime, where we once again observe that the massive spin-2 mode decreases the rate of mass-change and the rate of shift in the horizon radius. Our results indicate that intermediate-mass black holes (prime targets for  LISA) might be good candidates to test these aspects of modified gravity theories. Focusing on the $\ell=2, m=0$ modes, our analysis suggests that the backreaction effect may play a crucial role in the study of nonlinear memory of GWs in modified gravity theories. These are currently under investigation. 

The analysis in this work looks at the interesting possible signatures of strong gravity corrections in future GWs experiments. While the Starobinsky model and Stelle gravity are the low-energy quantum gravity action, these are not exhaustive. For instance, we have not included $R \ln \left(\Box R\right)$ and $R_{\mu\nu} \ln\left(\Box R^{\mu\nu}\right)$~\cite{Capozziello:2021krv}. It may be interesting to investigate the potential ring-down signatures of these terms as these terms may exceed the Stelle gravity contributions in low momenta. This is currently under investigation.

The decrease in measured energy flux and the decreased rate of horizon shift and mass change due to the massive spin-2 mode indicate the existence of quasi-bound states of the massive spin-2 modes surrounding the black hole~\cite{brito2016Thesis}. In the context of rotating black hole geometries, where this may lead to the formation of superradiantly induced spin-2 boson clouds, the question assumes greater significance. A detailed analysis of the massive spin-2 dynamics can resolve this question. However, such an analysis is beyond the scope of this paper so that we will leave it for future work. 
\section*{Acknowledgements}

The authors thank Susmita Jana,  S. Mahesh Chandran, and T. Parvez for the discussions. S.X. is financially supported by the MHRD fellowship at IIT Bombay. This work is supported by SERB-MATRICS grant.

\appendix
\section{Quadratic gravity as GR ``minus" Massive Gravity}\label{app:Msv-bigrav}
{
As first pointed out by Stelle~\cite{stelle1978GERG}, a generic quadratic theory of gravity is equivalent to GR ``minus" massive gravity at a linear level in flat space. In this section, we will demonstrate an extension of the analysis to a Ricci-flat spacetime and prove the consistency of the decomposition presented in Eqs.~\eqref{eq:ansatzRicciflat} and \eqref{eq:ansatz}. 
The generic quadratic gravity action in four dimensions is represented in Eq.~\eqref{eq:action} as,
{
\begin{equation}\label{eq:actionA}
\!\!\!\! S_{\rm QG}= \!\! \frac{1}{2\kappa^{2}} \int \!\! {d^{4}x \sqrt{-g}\left[R-2\kappa^{2}\alpha R_{\mu\nu}R^{\mu\nu}+2\kappa^{2}\beta R^{2}\right]}
\end{equation}
}
Perturbing the action in  \eqref{eq:actionA} about the background metric ($\bar g_{\mu\nu}$) to second order in $h_{\mu\nu}=g_{\mu\nu}-\bar g_{\mu\nu}$, we get,
\begin{equation}\label{eq:action-2}
\begin{split}
    S^{(2)}_{QG}[h_{\mu\nu}]= \frac{1}{2\kappa^{2}}\int d^{4}x \sqrt{-\bar g}\Bigl[-h^{\mu\nu}\delta R_{\mu\nu}[h_{\mu\nu}]-2 \alpha \kappa^2 \delta R_{\mu\nu}[h_{\mu\nu}] \delta R^{\mu\nu}[h_{\mu\nu}]+ 2 \beta \kappa^2 (\delta R[h_{\mu\nu}])^2
    +\frac{h}{2}\bar g^{\mu\nu}\delta R_{\mu\nu}\Bigr]
\end{split}
\end{equation}
where, $h=\bar g^{\mu\nu} h_{\mu\nu}$.
We now proceed by treating $\delta R_{\mu\nu}$ as an independent variable. However, this introduces extra degrees of freedom. 
In order to restrict the extra degrees of freedom, we replace 
$\delta R_{\mu\nu}$ in the above expression with an auxiliary variable $A_{\mu\nu}$, thus leading to: 
{\small
\begin{equation}\label{eq:action-2-aux}
 \begin{split}
   S^{(2)}_{QG}[-h_{\mu\nu},A_{\mu\nu}, \lambda_{\mu\nu}]= \frac{1}{2\kappa^{2}}\int d^{4}x \sqrt{-\bar g}\Bigl[-h^{\mu\nu} A_{\mu\nu}
   -2 \alpha \kappa^2 A_{\mu\nu} A^{\mu\nu}+ 2 \beta \kappa^2 A^2+\frac{h}{2}\bar g^{\mu\nu}A_{\mu\nu}
   +\lambda^{\mu\nu}(A_{\mu\nu}-\delta R_{\mu\nu}[h_{\mu\nu}])\Bigr]
    \end{split}
\end{equation}}
where $A=\bar{g}^{\mu\nu} A_{\mu\nu}$ and $\lambda_{\mu\nu}$ is the lagrange multiplier. Varying the above action with respect to $A_{\mu\nu}$, we get:
\begin{equation}\label{eq:constraint}
\lambda_{\mu\nu}= 4\alpha \kappa^2 A_{\mu\nu} - 4\beta \kappa^2 \bar g_{\mu\nu} A + h_{\mu\nu}-\frac{1}{2}h\bar g_{\mu\nu} ~.
\end{equation}
Rewriting the above expression in terms of $\lambda_{\mu\nu}$ and $h_{\mu\nu}$ and substituting the resultant in Eq.~\eqref{eq:action-2-aux}, we get: 

\begin{equation}
\begin{split}
    S^{(2)}_{QG}[\Phi_{\mu\nu},\tilde{\Phi}_{\mu\nu}]=\!\! \frac{1}{2\kappa^{2}}\int \!\! d^{4}x \sqrt{-\bar g}\Bigl[-\Phi^{\mu\nu}\delta R_{\mu\nu}\bigl[\Phi_{\mu\nu} \bigr]  + \tilde{\phi}^{\mu\nu}\delta R_{\mu\nu}\bigl[\tilde{\phi}_{\mu\nu}\bigr]+\frac{m^2}{2}\Bigl(\tilde{\Phi}_{\mu\nu} \tilde{\Phi}^{\mu\nu}-(1-\epsilon)\tilde{\Phi}^2\Bigr)\Bigr],
\end{split}
\end{equation}

where 
\begin{eqnarray}
\lambda_{\mu\nu}=\Phi_{\mu\nu}-\tilde{\Phi}_{\mu\nu};\\
\tilde{h}_{\mu\nu}=h_{\mu\nu}-\frac{1}{2} h \bar{g}_{\mu\nu}=\Phi_{\mu\nu}+\tilde{\Phi}_{\mu\nu}\\
\epsilon = \frac{1+\beta}{(\alpha-4\beta)};\quad m^2= \frac{1}{(\alpha \kappa^2)}    
\end{eqnarray}

In the case of GUP-inspired Quadratic gravity, Eq.~\eqref{eq:constraint} reduces to:
\begin{equation}
    h_{\mu\nu} = \psi_{\mu\nu}-\frac{1}{2}\bar g_{\mu\nu} \psi-4 \gamma \hat{A}_{\mu\nu}-\gamma A \bar{g}_{\mu\nu}
\end{equation}
where, $\hat{A}_{\mu\nu}= A_{\mu\nu}-\frac{1}{4}\bar g_{\mu\nu} A$  is the traceless part of $A_{\mu\nu}$.
Mapping $\lambda_{\mu\nu} \to \psi_{\mu\nu}$,  $\hat{A}_{\mu\nu} \to \hat{R}^{(1)}_{\mu\nu}$  and $A\to R^{(1)}$, leads to  Eqs.~(\ref{eq:ansatzRicciflat},\ref{eq:ansatz}). The above analysis establishes the metric decomposition into massless spin-2, massive spin-0, and massive spin-2 modes. Combining the results of Appendix 
\eqref{app:EOM-generic}, this shows the equations of motion (\ref{eq:Rflatboxpsi}, \ref{eq:RflatboxRic}, \ref{eq:lintrcEOM_flat}) leads to the correct equations of motion.

It is possible to redo the above analysis for an arbitrary Einsteinian manifold, interms of the Einstein tensor instead of the Ricci tensor ~\cite{tachinami2021PRD}. Following Ref.~\cite{tachinami2021PRD}, we can represent the quadratic gravity action \eqref{eq:action} as:
\begin{equation}\label{eq:actionAG}
\!\!\!\! S_{\rm QG}= \!\! \frac{1}{2\kappa^{2}} \int \!\! {d^{4}x \sqrt{-g}\left[-G-2\kappa^{2}\alpha G_{\mu\nu}G^{\mu\nu}+2\kappa^{2}\beta G^{2}\right]} \, .
\end{equation}
Perturbing the action \eqref{eq:actionAG} about the background metric ($\bar g_{\mu\nu}$) to second order in $h_{\mu\nu}=g_{\mu\nu}-\bar g_{\mu\nu}$, we get,
\begin{equation}\label{eq:action-2G}
\begin{split}
    S^{(2)}_{QG}[h_{\mu\nu}]= \frac{1}{2\kappa^{2}}\int  d^{4}x \sqrt{-\bar g}\Bigl[h^{\mu\nu}\delta G_{\mu\nu}[h_{\mu\nu}]
    -2 \alpha \kappa^2 \delta G_{\mu\nu}[h_{\mu\nu}] \delta G^{\mu\nu}[h_{\mu\nu}]
    + 2 \beta \kappa^2 (\delta G[h_{\mu\nu}])^2
    -\frac{h}{2}\bar g^{\mu\nu}\delta G_{\mu\nu}\Bigr]
    \end{split}
\end{equation}
where, $h=\bar g^{\mu\nu} h_{\mu\nu}$.
Replacing $\delta G_{\mu\nu}$ in the above expression by an auxiliary variable $A_{\mu\nu}$ and imposing a constraint, we get,
{\small
\begin{equation}\label{eq:action-2-aux2}
\begin{split}
  S^{(2)}_{QG}[h_{\mu\nu},A_{\mu\nu}, \lambda_{\mu\nu}]= \frac{1}{2\kappa^{2}}\int d^{4}x \sqrt{-\bar g}\Bigl[-h^{\mu\nu} A_{\mu\nu}
  -2 \alpha \kappa^2 A_{\mu\nu} A^{\mu\nu}+ 2 \beta \kappa^2 A^2
  -\frac{h}{2}\bar g^{\mu\nu}A_{\mu\nu}
  +\lambda^{\mu\nu}(A_{\mu\nu}-\delta G_{\mu\nu}[h_{\mu\nu}])\Bigr]
  \end{split}
\end{equation}
}
where $A=\bar{g}^{\mu\nu} A_{\mu\nu}$ and $\lambda_{\mu\nu}$ is the lagrange multiplier. Varying the above action with respect to $A_{\mu\nu}$, we get
\begin{equation}\label{eq:constraint2}
\lambda_{\mu\nu}= 4\alpha \kappa^2 A_{\mu\nu} - 4\beta \kappa^2 \bar g_{\mu\nu} A - h_{\mu\nu}+\frac{1}{2}h\bar g_{\mu\nu} ~.
\end{equation}
Substituting $A_{\mu\nu}$ in terms of $\lambda_{\mu\nu}$ and $h_{\mu\nu}$ from Eq.~\eqref{eq:constraint2} in Eq.~\eqref{eq:action-2-aux2} and using
\begin{equation}
    \tilde{h}_{\mu\nu}=h_{\mu\nu}-\frac{1}{2} h \bar{g}_{\mu\nu}=\Phi_{\mu\nu}+\tilde{\Phi}_{\mu\nu}~; \quad
    \lambda_{\mu\nu}=\tilde\Phi_{\mu\nu}-{\Phi}_{\mu\nu},
\end{equation}
we get,
\begin{equation}
\begin{split}
S^{(2)}_{QG}[\Phi_{\mu\nu},\tilde{\Phi}_{\mu\nu}]=\frac{1}{2\kappa^{2}}\int \!\! d^{4}x \sqrt{-\bar g}\Bigl[\Phi^{\mu\nu}\delta G_{\mu\nu}\bigl[\Phi_{\mu\nu} \bigr] 
- \tilde{\Phi}^{\mu\nu}\delta G_{\mu\nu}\bigl[\tilde{\Phi}_{\mu\nu}\bigr]
+\frac{m^2}{2}\Bigl(\tilde{\phi}_{\mu\nu} \tilde{\Phi}^{\mu\nu}-(1-\epsilon)\tilde{\Phi}^2\Bigr)\Bigr],
\end{split}
\end{equation}
where, $\epsilon=1+\beta/(\alpha-4\beta)$,  $m^2=1/(\alpha \kappa^2)$. 
For the GUP-inspired Stelle gravity, we once again recover the ansatz \eqref{eq:ansatzRicciflat} with the following mapping $\psi_{\mu\nu}\to-\lambda_{\mu\nu}$~,  $\hat{R}^{(1)}_{\mu\nu}\to-\hat{A}_{\mu\nu}$ and $R^{(1)}\to-A$. 
}
\section{Reduced equations of motion in Ricci-flat space-times}\label{app:EOM-generic}
The field equation derived from the action~\eqref{eq:action} with arbitrary values of the parameters $\alpha$ and $\beta$ is given by
\begin{equation}\label{eq:EOM-effT-genA}
    \begin{split}
&R_{\mu\nu}-\frac{1}{2} R g_{\mu\nu}+4 \beta R_{\mu\nu} R - 4 \alpha R^{\rho \sigma} R_{\mu \rho \nu \sigma} + 2 (\alpha - 2\beta) \nabla_\mu \nabla_\nu R \\
&- 2 \alpha \Box R_{\mu \nu}+ g_{\mu \nu}\left( \alpha R_{\rho \sigma} R^{\rho \sigma} +\beta R^2 + (\alpha -4 \beta) \Box R \right)=0~.
\end{split}
\end{equation}
Linearizing the above field equation in a Ricci-flat background one gets \begin{equation}\label{eq:linEOMTeff-genA}
\begin{split}
        \delta R_{\mu\nu}-\frac{1}{2}\bar g_{\mu\nu} \delta R- 4 \alpha \bar R_{\mu\rho\nu\sigma} \delta R^{\rho\sigma}\\+ 2 (\alpha - 2\beta) \bar\nabla_\mu  \bar\nabla_\nu \delta R- 2 \alpha \bar \Box \delta R_{\mu\nu}\\ - \bar g_{\mu \nu} \left( \alpha-4 \beta \right) \bar\Box \delta R = 0 \, .
\end{split}
\end{equation}
Equation~\eqref{eq:linEOMTeff-genA} can be written in terms of the metric perturbations by expanding $\delta R_{\mu\nu}$ and $\delta R$ in terms of $h_{\mu\nu}$ (see Eqs.~(\ref{eq:ricci},\ref{eq:ricciscalar}) and Ref.~\cite{MTW}). We choose the following ansatz for the metric perturbations in Ricci-flat spacetime:
\begin{equation}\label{eq:ansatz-genA}
\!\!\! h_{\mu\nu}=\left[\psi_{\mu \nu}- \frac{\bar g_{\mu\nu}}{2} \psi \right]+ \left[C_1+ \frac{C_2}{4}\right] R^{(1)} \bar g_{\mu\nu} - C_2 \hat{R}^{(1)}_{\mu\nu}~,
\end{equation}
where $C_1, C_2$ are constants to be determined and $\psi_{\mu\nu }$ satisfies the transverse-traceless condition $\left(\bar\nabla^\mu \psi_{\mu\nu}=0, ~\bar{g}^{\mu\nu} \psi_{\mu\nu}=0\right)$.
We demand that $R^{(1)}$ and $\hat{R}^{(1)}_{\mu\nu}$ to be respectively equal to the perturbed Ricci scalar $\delta R$ and the \emph{traceless part} of the perturbed Ricci tensor ($\delta R_{\mu\nu}-\frac{1}{4} \bar g_{\mu\nu}\delta R$). This will ensure that the linearized Bianchi identity \eqref{eq:pertbian} is satisfied  at all times and in all spaces. Hence, equating $\delta R$ with $R^{(1)}$ and comparing it with the trace of linearized field equation~\eqref{eq:linEOMTeff-genA} we get,
\begin{equation}\label{eq:C2}
C_2=-3C_1 +4\left(\alpha-3 \beta\right)~.
\end{equation}
Thus, we get the propagation equation for the massive spin-0 mode as,
\begin{equation}\label{eq:EOM-spin0-genA}
\bar{\Box}R^{(1)} -  R^{(1)}/{4(3\beta - \alpha)}=0~. 
\end{equation}
Similarly, equating $R^{(1)}_{\mu\nu}=\hat{R}^{(1)}_{\mu\nu}+\frac{1}{4}\bar g_{\mu\nu} R^{(1)}$ with $\delta R_{\mu\nu}$ and using Eq.~\eqref{eq:C2} we get,
\begin{equation}\label{eq:ricci-genA}
\begin{split}
&{R}^{(1)}_{\mu\nu}+\frac{1}{2}\Bigl[\Bigl(\bar\Box\psi_{\mu\nu} + 2 \bar R_{\mu\rho\nu\sigma}\psi^{\rho\sigma}\Bigr)
-\left(C_1 - 4 \alpha+12\beta\right)\Bigl(\bar\nabla_\mu \bar\nabla_\nu R^{(1)}-\frac{\bar g_{\mu\nu} R^{(1)}}{8(\alpha-3\beta)}\Bigr)+\\
&\left(3C_1-4\alpha+12\beta\right)\Bigl(\bar\Box {R}^{(1)}_{\mu\nu}+2 \bar R_{\mu\rho\nu\sigma} {R}^{(1) \rho \sigma}\Bigr)\Bigr]=0
\end{split}
\end{equation}
Comparing with Eq.~\eqref{eq:linEOMTeff-genA} and using Eq.~\eqref{eq:lintrcEOM_flatb} we get,
\begin{eqnarray}
 \bar{\Box}\psi_{\mu\nu}+ 2 \bar{R}_{\mu\alpha\nu\beta} \psi^{\alpha\beta} = 0~,\label{eq:EOM-graviton-genA}\\
 C_1= -4 \beta \mbox{ and } \alpha=2\beta~.\label{eq:C1}
\end{eqnarray}
Equation~\eqref{eq:EOM-graviton-genA} gives the dynamics of the massless spin-2 graviton mode, as $\psi_{\mu\nu}$ satisfies the traceless-transverse condition. Equation~\eqref{eq:C1}, on the other hand, implies that the decomposition scheme in Eq.~\eqref{eq:ansatz-genA} is specific to the GUP-inspired Stelle gravity $(\alpha=2\beta=\gamma)$, in which case $C_1=-4\beta=-2\gamma$ and $C_2=4\gamma$. Thus, substituting $C_1$ and  $C_2$ in Eq.~\eqref{eq:ricci-genA} and taking cognizance of Eqs.~\eqref{eq:EOM-spin0-genA}, \eqref{eq:EOM-graviton-genA}, \eqref{eq:pertbian}  and \eqref{eq:spin-2tracls}, we get the propagation equation of the massive spin-2 mode as given in Eq.~\eqref{eq:RflatboxRic}
\begin{equation}
\bar{\Box}\hat{R}^{(1)}_{\mu \nu} +2 \bar{R}_{\mu\alpha \nu\beta} \hat{R}^{(1) \alpha \beta} - \hat{R}^{(1)}_{\mu \nu}/{(2 \gamma)} =0 \, .
\end{equation}
In the case of $\alpha=2\beta=\gamma$, Eq.~\eqref{eq:EOM-spin0-genA} reduces to Eq.~\eqref{eq:lintrcEOM_flat}. It is important to note that in the massless limit $\gamma\rightarrow 0$, Eqs.~\eqref{eq:lintrcEOM_flat} and \eqref{eq:RflatboxRic} suggest that $\hat{R}^{(1)}_{\mu\nu}$ and $R^{(1)}$ vanishes identically and the theory is described by only one massless spin-2 graviton $\psi_{\mu\nu}$.
\onecolumngrid
\section{Coefficients in the effective GW stress-energy tensor}\label{app:GWSET}

Using \href{http://www.xact.es/}{xAct Mathematica packages}, $\mathcal{A}_{\mu\nu},\mathcal{B}_{\mu\nu},\mathcal{C}_{\mu\nu}$, and $\mathcal{D}_{\mu\nu}$ in Eq. \eqref{eq:teff_full_trcls} are obtained as:
\flushleft
\begin{align}
\mathcal{A}_{\mu \nu}&=-\bar R_{\mu \lambda \alpha \rho } \psi_{\nu }^{\alpha } \psi^{\lambda \rho } - \frac{1}{4} \bar{g}_{\mu \nu }\bar  R_{\alpha \rho \lambda \sigma \
} \psi^{\alpha \lambda } \psi^{\rho \sigma }  \\
\label{app:eq2}
\mathcal{B}_{\mu \nu}&=
2 \bar R_{\nu }{}^{\sigma }_{\alpha }{}^{\eta } \bar R_{\lambda \sigma \rho \eta } \psi_{\mu }^{\alpha } \psi^{\lambda \rho } - 4 \bar R_{\mu \alpha \lambda }{}^{\eta } \bar R_{\nu \rho \sigma \eta } \psi^{\alpha \lambda } \psi^{\rho \sigma } - 2 \bar R_{\mu \alpha \nu }{}^{\eta } \bar R_{\lambda \rho \sigma \eta } \psi^{\alpha \lambda } \psi^{\rho \sigma} 
- \bar R_{\mu \rho \nu \sigma } \bar\nabla^{\rho }\psi^{\alpha \lambda } \bar\nabla^{\sigma }\psi_{\alpha \lambda } \nonumber   \\
&
+ 4 \bar R_{\mu}{}^{\eta }{}_{\nu \alpha } \bar R_{\lambda \rho \sigma \eta } \psi^{\alpha \lambda } \psi^{\rho \sigma } 
- \bar{g}_{\mu \nu } \bar R_{\alpha }{}^{\eta }{}_{\lambda }{}^{h} \bar R_{\rho \eta \sigma h} \psi^{\alpha \lambda } \psi^{\rho \sigma }
+ 2 \bar R_{\alpha \rho \lambda \sigma } \bar\nabla_{\mu }\psi^{\alpha \lambda } \bar\nabla_{\nu }\psi^{\rho \sigma } - 2 \bar R_{\alpha \rho \lambda \sigma } \bar\nabla_{\mu }\psi^{\rho \sigma } \bar\nabla^{\lambda }\psi_{\nu }{}^{\alpha } 
\nonumber   \\
& + 2 \bar R_{\alpha \lambda \rho \sigma } \bar\nabla_{\nu }\psi_{\mu }{}^{\alpha } \bar\nabla^{\sigma }\psi^{\lambda \rho } - 2 \bar R_{\alpha \lambda \rho \sigma } \bar\nabla^{\alpha }\psi_{\mu \nu } \bar\nabla^{\sigma }\psi^{\lambda \rho} \\
\label{app:eq3}
\mathcal{C}_{\mu \nu}&=
    8 \bar R_{\nu \lambda \alpha \rho } \hat{R}^{(1)}{}_{\mu }{}^{\alpha } \hat{R}^{(1)}{}^{\lambda \rho } + 2 \bar R_{\mu \lambda \nu \rho } (4 \hat{R}^{(1)}{}_{\alpha }{}^{\rho } \hat{R}^{(1)}{}^{\alpha \lambda } - 7 R^{(1)} \hat{R}^{(1)}{}^{\lambda \rho })  - 8 \bar R_{\alpha \lambda \rho \sigma } \bar\nabla_{\nu }\hat{R}^{(1)}{}_{\mu }{}^{\alpha } \bar\nabla^{\sigma }\psi^{\lambda \rho }  
    \nonumber  \\
        &
     + 4 \bar{g}_{\mu \nu } R_{\alpha \rho \lambda \sigma } \hat{R}^{(1)}{}^{\alpha \lambda } \hat{R}^{(1)}{}^{\rho \sigma } 
          - 16 \bar R_{\alpha \rho \lambda \sigma } \bar\nabla_{\mu }\psi^{\rho \sigma } \bar\nabla_{\nu }\hat{R}^{(1)}{}^{\alpha \lambda } + 2 \bar R_{\nu \rho \lambda \sigma } \bar\nabla_{\mu }\psi^{\rho \sigma } \bar\nabla^{\lambda}R^{(1)} 
        \nonumber   \\
        &
          + 8 \bar R_{\alpha \rho \lambda \sigma } \bar\nabla_{\mu }\psi^{\rho \sigma } \bar\nabla^{\lambda }\hat{R}^{(1)}{}_{\nu }{}^{\alpha } 
          + 2 \bar R_{\nu \lambda \rho \sigma } \bar\nabla^{\lambda}R^{(1)} \bar\nabla^{\sigma }\psi_{\mu }{}^{\rho} + 2 \bar R_{\nu \sigma \lambda \rho } \bar\nabla^{\lambda }R^{(1)} \bar\nabla^{\sigma }\psi_{\mu }{}^{\rho } 
     \nonumber \\ &
     + 4 \bar R_{\mu \rho \nu \sigma } \bar\nabla^{\rho }\hat{R}^{(1)}{}^{\alpha \lambda } \bar\nabla^{\sigma }\psi_{\alpha \lambda } 
          + 4 \bar R_{\mu \sigma \nu \rho } \bar\nabla^{\rho }\hat{R}^{(1)}{}^{\alpha \lambda } \bar\nabla^{\sigma }\psi_{\alpha \lambda } - 10 \bar R_{\mu \lambda \rho \sigma } \bar\nabla_{\nu }R^{(1)} \bar\nabla^{\sigma }\psi^{\lambda \rho } 
      \nonumber     \\
        &
        + 8 \bar R_{\alpha \lambda \rho \sigma } \bar\nabla^{\alpha }\hat{R}^{(1)}{}_{\mu \nu } \bar\nabla^{\sigma }\psi^{\lambda \rho } + 2 \bar{g}_{\mu \nu } \bar R_{\lambda \rho \sigma \eta } \bar\nabla^{\lambda }R^{(1)} \bar\nabla^{\eta }\psi^{\rho \sigma }\\
 %
\mathcal{D}_{\mu \nu}&=
     4 \Big(8 \bar R_{\nu }{}^{\sigma }{}_{\alpha }{}^{\eta } \bar R_{\lambda \sigma \rho \eta } \hat{R}^{(1)}{}_{\mu }{}^{\alpha } \hat{R}^{(1)}{}^{\lambda \rho } - 16 \bar R_{\mu \alpha \lambda }{}^{\eta } \bar R_{\nu \rho \sigma \eta } \hat{R}^{(1)}{}^{\alpha \lambda } \hat{R}^{(1)}{}^{\rho \sigma } 
      + 2 \bar R_{\nu \sigma \lambda \rho } \bar\nabla^{\lambda }R^{(1)} \bar\nabla^{\sigma }\hat{R}^{(1)}{}_{\mu }{}^{\rho} \nonumber    \\
      &
     - 8 \bar R_{\mu \alpha \nu }{}^{\eta } \bar R_{\lambda \rho \sigma \eta } \hat{R}^{(1)}{}^{\alpha \lambda } \hat{R}^{(1)}{}^{\rho \sigma } + 16 \bar R_{\mu }{}^{\eta }{}_{\nu \alpha } \bar R_{\lambda \rho \sigma \eta } \hat{R}^{(1)}{}^{\alpha \lambda } \hat{R}^{(1)}{}^{\rho \sigma } 
   \nonumber        \\
       &
     - 4 \bar{g}_{\mu \nu } \bar R_{\alpha }{}^{\eta }{}_{\lambda }{}^{h} \bar R_{\rho \eta \sigma h} \hat{R}^{(1)}{}^{\alpha \lambda } \hat{R}^{(1)}{}^{\rho \sigma } + 8 \bar R_{\alpha \rho \lambda \sigma } \bar\nabla_{\mu }\hat{R}^{(1)}{}^{\alpha \lambda } \bar\nabla_{\nu }\hat{R}^{(1)}{}^{\rho \sigma }   \\
   &
    + \bar R_{\mu \lambda \nu \sigma } \bar\nabla^{\lambda }R^{(1)} \bar\nabla^{\sigma }R^{(1)}+ 2 \bar R_{\nu \lambda \rho \sigma } \bar\nabla^{\lambda }R^{(1)} \bar\nabla^{\sigma }\hat{R}^{(1)}{}_{\mu }{}^{\rho }
- 4 \bar R_{\mu \rho \nu \sigma } \bar\nabla^{\rho }\hat{R}^{(1)}{}^{\alpha \lambda } \bar\nabla^{\sigma }\hat{R}^{(1)}{}_{\alpha \lambda }\Big)  \nonumber 
\end{align}
%

As mentioned earlier, we assume $\psi(V,\rho)$ and $P(V,\rho)$ to be slowly varying close to the event horizon $(\rho\sim2M_0)$. The leading order solid angle-averaged value (of the $\rho-V$ component) of the effective GW stress-energy tensor (up to leading order corrections in $\gamma$) used in Eq.~\eqref{eq:JP} in the Schwarzschild background is given by
\begin{align}
\!\!\! t_V^{\rho\rm GW}\approx
    & \frac{M_{0}^4 \psi' (V,\rho) }{21 \pi}\left\{8[21o^{00}\iota^{22}+20o^{23}\iota^{23}+8o^{33}\iota^{33}]\gamma P' (V,\rho)+[21( o^{22})^2+20(o^{23})^2+8(o^{33})^2]\psi' (V,\rho)  \right\} \nonumber
   \\&
    + \frac{1}{21 \pi} M_{0}^{2} \psi^{\prime} (V,\rho)\left[-4(21 o^{12} \iota^{02}+10 o^{13} \iota^{03}+21 o^{02} \iota^{12}+10 o^{03} \iota^{13}) \gamma P^{\prime} (V,\rho)+\right. \nonumber \\
    &\left.\left(21 o^{02} o^{12} +10 o^{03} o^{13} +\gamma\left(21 (o^{22})^{2}-20 (o^{23})^{2}+40 o^{22} \iota^{33}+8 (o^{33})^{2}\right) \right) \psi^{\prime} (V,\rho)\right]
\nonumber \\&
    +\frac{1}{21 \pi}\left(21 o^{12} \iota^{02}+10 o^{13} \iota^{03}-21 o^{02} \iota^{12}-10 o^{03} \iota^{13}\right) M_{0}  \gamma\left[\psi (V,\rho) \mathrm{P}^{\prime} (V,\rho)-\mathrm{P} (V,\rho) \psi^{\prime} (V,\rho)\right]
 \nonumber   \\&
  +  \frac{1}{672 M_{0}^2\pi}     
    \left[84(o^{11} \iota^{00}+o^{00} \iota^{11} )  \gamma P (V,\rho) \psi (V,\rho)+\left(-21o^{00}o^{11} +2\left(-42 o^{02} o^{12}+21 (o^{12})^{2}
     \right.\right.\right. \nonumber \\&\left.\left.\left.
    -20 o^{03} o^{13}+10 (o^{13})^{2}-63 o^{11} o^{22}+38 o^{11} o^{33}\right) \gamma \right) (\psi (V,\rho))^{2}+84(o^{01})^2\gamma(\psi' (V,\rho))^{2} \right]
\nonumber    \\&
    -\frac{1}{32 M_{0}^{3} \pi }\left(2(o^{01})^2+o^{00}o^{11}-2o^{01}  o^{11}\right) \gamma \psi (V,\rho)\psi' (V,\rho) +\frac{1}{64 M_{0}^{4} \pi }  (\psi (V,\rho))^2\gamma \left[-2(o^{01})^2+o^{00}  o^{11}\right]
\nonumber \\&
    +\frac{\gamma}{168\pi}\left[42(-o^{11}\iota^{00}+o^{00}\iota^{11} ) P (V,\rho)\psi^{\prime} (V,\rho)
    +\psi (V,\rho)\left( 42(o^{11} \iota^{00}-o^{00} \iota^{11} )  P^{\prime} (V,\rho)+\left( 21 o^{02} o^{12}
    \right.\right.\right.
\nonumber  \\&
    \left.\left.\left.
    -21 (o^{12})^{2}+10o^{03} o^{13}-10 (o^{13})^{2}-42 o^{11} o^{22}-20 o^{11} o^{33}\right) \psi^{\prime} (V,\rho)\right)\right]
\nonumber \\&
    +\frac{1}{336 \pi}\left[4(21 o^{12} \iota^{02}+10 o^{13} \iota^{03}+21 o^{02} \iota^{12}+10 o^{03} \iota^{13} )   \gamma P (V,\rho) \psi (V,\rho)\right. 
 \nonumber    \\
    &\left.-\left(21 o^{02} o^{12}+42 (o^{12})^{2}+10 o^{03} o^{13}+20 (o^{13})^{2}-42 o^{11} o^{22}-20o^{11} \iota^{33}\right)  \psi (V,\rho)^{2}
    \right. 
    \nonumber \\&
\left. +2 \psi^{\prime} (V,\rho)\left\{-84(o^{11} \iota^{00}+2 o^{01} \iota^{01}+o^{00} \iota^{11} )  \gamma P^{\prime} (V,\rho)+\left(21 (o^{01})^{2}  \right.\right.\right.
    \nonumber \\&\left.\left.\left.
    +21 o^{00} o^{11} +252 o^{02} o^{12} \gamma+120 o^{03} o^{13} \gamma-84o^{01}o^{22} \gamma-40o^{01} \iota^{33} \gamma\right) \psi^{\prime} (V,\rho)\right\}\right]    
\end{align}

where $'$ denotes derivative wrt $V$.

\vspace{1cm}
\def\bibsection{}  
\centerline{\small {\textbf{REFERENCES}} }
\bigskip
\bibliography{stelle_ref} 
\end{document}